**Title**

# Fabrication and characterization of vacuum deposited fluorescein thin films

**Authors**

Pasi Jalkanen[1,2], Sampo Kulju[1], Konstantin Arutyunov[1], Liisa Antila[3], Pasi Myllyperkiö[3], Jouko Korppi-Tommola[3], Teemu Ihalainen[4], Tommi Kääriäinen[5], Marja-Leena Kääriäinen[5]

*Corresponding author*

*Konstantin Arutyunov*, konstantin.arutyunov@jyu.fi,
Jyväskylä University, Department of Physics, Nanoscience center (NSC)
PL 35
FI-40014 Jyväskylä,
FINLAND

Tel: +358 14 260 4769
Fax: +358 14 260 4756

**Affiliation**
(1) University of Jyväskylä, Department of Physics, Nanoscience center (NSC)
P.O.Box 35, FI-40014 Jyväskylä, FINLAND.
(2) AALTO University, Department of Materials Science and Engineering, Nordic Hysitron Laboratory, P.O.Box 16200, FI-00076, Espoo, FINLAND
(3) University of Jyväskylä, Department of Chemistry, Nanoscience center (NSC)
P.O.Box 35, FI-40014 Jyväskylä, FINLAND
(4) University of Jyväskylä, Department of Biology, Nanoscience center (NSC)
P.O.Box 35, FI-40014 Jyväskylä, FINLAND
(5) Lappeenranta University of Technology, ASTRaI
P.O.Box 181 FI-50101 Mikkeli, FINLAND

**Email**

P. Jalkanen:           pasi.jalkanen@gmail.com
S. Kulju:              sampo.j.kulju@jyu.fi
K. Arutyunov:          konstantin.arutyunov@jyu.fi
L. Antila:             liisa.j.antila@jyu.fi
P. Myllyperkiö:        pasi.myllyperkio@jyu.fi
J. Korppi-Tommola:     jouko.korppi-tommola@jyu.fi
T. Ihalainen:          teemu.o.ihalainen@jyu.fi
T. Kääriäinen:         tommi.kaariainen@lut.fi
M-L. Kääriäinen:       marja-leena.kaariainen@lut.fi

**Abstract**

Simple vacuum evaporation technique for deposition of dyes on various solid surfaces has been developed. The method is compatible with conventional solvent-free nanofabrication processing enabling fabrication of nanoscale optoelectronic devices. Thin films of fluorescein were deposited on glass, fluorine-tin-oxide (FTO) coated glass with and without atomically layer deposited (ALD) nanocrystalline 20 nm thick anatase $TiO_2$ coating. Surface topology, absorption and emission spectra of the films depends on their thickness and the material of supporting substrate. On a smooth glass surface the dye initially formes islands before merging into a uniform layer after 5 to 10 monolayers. On FTO covered glass the absorption spectra are similar to fluorescein solution in ethanol. Absorption spectra on ALD-$TiO_2$ is red shifted compared to the film deposited on bare FTO. The corresponding emission spectra at λ = 458 nm excitation show various thickness and substrate dependent features, while the emission of films deposited on $TiO_2$ is quenched due to the effective electron transfer to the semiconductor conduction band.

## 1. Introduction

Molecular engineering can be applied to modify the surface states of metals and semiconductors [1]. Dye molecules, widely utilized in cell biology and dye sensitized solar cells (DSSC) [2-3], provide high efficiency of photon to electron conversion making them a promising material for various electro-optical applications [1]. With the rapid development of nano- electronics and photonics there is an urge for the new class of systems containing functionalized molecular layers on dry surfaces [5]. Development of a vacuum-compatible solvent-free dye deposition technology is highly desirable [6].

Various "macroscopic" techniques - transmission, absorption, fluorescence spectroscopy and confocal microscopy – have been successfully used to study properties of dyes in various solutions. One of the most promising applications – DSSC - explicitly utilizes the matching between the dye's LUMO level and the bottom of the conduction band of the semiconducting electrode (e.g. $TiO_2$). It is well known that molecular properties are subject to change on aggregation and/or in contact with a semiconductor or metal surface [7]. Modern nanotechnology is capable to fabricate sub-10 nm structures [8,9] enabling resolution of very fine details of the energy spectra using solid state tunneling technique [10]. The method is expected to be applicable to various systems, including organic dyes. However, it requires solvent-free "clean" fabrication of samples with the molecular-size objects attached to solid surfaces. The development of the corresponding technique and analysis of the physical properties of the fabricated fluorescein thin films is the main topic of the paper.

Several vacuum evaporation methods have been suggested for fabrication of molecular thin films [6,11-13]. Widely used "wet" deposition of dyes from a solution typically leads to monolayer formation where one molecule occupies 1 to 2 nm$^2$ surface area. In the case of vacuum evaporation the physical properties of the grown thin film can be varied by controlling the vacuum conditions, substrate temperature, vapor flux and kinetic energy of the molecules [11, 14]. All these factors might appear crucial for the performance of a device containing molecule-semiconductor or molecule-metal interface(s). For example, if the deposited dye molecules strongly aggregate, then the electron injection or fluorescence may be considerably suppressed due to the molecule-molecule energy transfer.

## 2. Experimental Details

For our experiments we utilized the fluorescein dye (*Merck Inc.* No. 3990). The dye powder was heated in a carbon crucible by the electric heater (Fig. 1). Temperature of the crucible was controlled with Pt-100 film sensor. One hour pre-heating with ~ 8 W power was found to be sufficient for stabilizing the temperature of the dye. During this time the evaporated molecules were pumped from the evaporation chamber via bypass valve, and the access to the condensation (target) chamber was blocked by the shutter. *Pfeiffer TH-260* turbo pump was connected directly to the condensation chamber being back-pumped through the activated alumina filter by *Alcatel Pascal 2010* rough pump providing oil-free vacuum conditions. The circular aperture between the evaporation (30 cm$^3$) and the condensation (~0.5 L) chambers had a diameter 8 mm being located at a distance 40 mm from the target. After the pre-heating, the shutter was opened and the evaporating molecules were allowed to enter the condensation chamber. Target holder was attached to the cooled chamber lid kept at temperature 24 °C - 30 °C. Evaporation rate and the layer thickness were monitored by the crystal oscillator with *Leybold Inficon Deposition* controller setup. The dye deposition rate could be varied by changing the heating power and/or the ambient pressure. At typical parameters (pressure 1.5x10$^{-4}$ Pa and the crucible temperature from 150 °C to 160 °C) the deposition rate could be stabilized in the range from 0.1 Å/s to 2.0 Å/s. According to the atomic force microscope (AFM) analysis (Veeco Digital instruments NanoScope IV, tapping mode, PPP-NCH and SSS-NCH tips with the tip radius < 10 nm and < 5 nm, respectively), the evaporated film thickness varied ±17 % from the center to the edge of the circular film with diameter ~ 1 cm.

Titanium dioxide thin films were deposited on FTO (F:SnO$_2$) substrates using atomic layer deposition (ALD) technique (*TFS-500 ALD* reactor by *Beneq Oy*, Vantaa, Finland). The structure of the films was anatase and they were grown at 300 °C. Titanium tetrachloride (99.0 % *Fluka*) and ion exchanged water were used as precursors. Nitrogen (99.999 %, *AGA*) was used as a carrier and purging gas. The reactor operated at a pressure from 5×10$^2$ Pa to 1×10$^3$ Pa. The precursors were kept at 20 °C during the deposition. The thickness of the films was 20 nm. The surface of FTO-covered was found to have root mean square (RMS) roughness ranging from 10 nm to 20 nm. Thin microscope cover glass with surface RMS roughness < 1 nm was utilized as an alternative substrate enabling reliable AFM analysis of the deposited fluorescein films.

**Figure 1.** The evaporator setup. The sample and the crystal oscillator for thickness measurement were assembled symmetrically at the same distance from the crucible.

Properties of dye functionality on solid surfaces can be studied using absorption spectroscopy and confocal microscopy measurements. To some extent the quality of the fabricated dye films can be judged by the optical absorption. Confocal microscope gives information on the lateral distribution of the optical emission. For example, on glass surface no electron injection occurs, and therefore, fluorescence is expected to be clearly detectable. On the contrary, on $TiO_2$ surface no fluorescence should be observed due to the effective electron injection to the conduction band of the semiconductor [15,16]. Recently it was reported that in N719 dye deposited on $TiO_2$ the surface charge diminishes under illumination [13].

Absorption spectra were measured with *Perkin-Elmers LAMBDA 850* spectrophotometer with slit width 2 nm and the wavelength between λ = 400 nm and λ = 800 nm. Dye emission spectra were obtained by confocal microscope *Olympus Fluoview 1000* by applying excitation wavelengths λ = 458 nm and by using *UPLSAPO* 20x (NA = 0.7) objective. The slit width was adjusted to 10 nm and the spectra were collected with a step size of 5 nm. Thickness of the films was measured by *Tencor P-10* profilometer and *Dimension 3100* atomic force microscope (AFM).

## 3. Results

Typical profiles of the edges of the two dye films of various thickness deposited through a mask onto glass are shown in Fig. 2. On the substrate within the locus of the film edge one can clearly observe ~ 10 nm high formations, which are absent on the surface of the deposited fluorescein film few μm away from the edge. We associate the effect with the initial film growth described by the island formation mechanism, while the layer-plus-island mode cannot be excluded either [11, 14]. Difference of the surface roughness of the thin (Fig. 2a) and the relatively thick (Fig.2b) samples also supports the island formation scenario when the initial inhomogeneity is healed with the increase of the film thickness. The island formation mode suggests that molecular interactions on the surface are dominant. A different behavior is expected for films deposited onto $TiO_2$ surface. The binding of fluorescein on $TiO_2$ surface (mono-molecular layer) most probably takes place via covalent bonding of the carboxylic group to titanium atoms [15, 17]. Therefore, layer-by-layer or layer-plus-island mode is expected once the initial monolayer is formed, the further film growth is controlled by the van der-Waals interactions. After a certain thickness threshold is reached the subsequent layers are generated by molecular aggregation described by the layer-by-layer mode independently of the properties of the substrate. Unfortunately, much higher initial roughness of the FTO covered glass, with and without ALD-deposited $TiO_2$, makes the quantitative comparison of the deposited dye films with the ones grown on glass substrates not representative. However, in both cases the absence of pronounced narrow peaks or dips suggests the layer-by-layer subsequent scenario. It is expected that the molecules on TiO$_2$ surface should be distributed more evenly forming a continuous film at earlier stages of the growth compared to glass substrates.

**Figure 2.** Typical cross sections of the two fluorescein film edges of different thickness on glass substrate measured by atomic force microscope. Insets show the corresponding quasi-3D plots.

### 3.1 Absorption spectra

To examine the absorption properties of the vacuum-deposited films the solution spectra shown in Fig. 3 were taken as a reference. The absorption spectra, after subtraction of the background, for the 30 nm to 90 nm thick dye films are depicted in the same figure. Absorption spectra for the 30 nm thick fluorescein film on FTO covered glass shows a blue shifted band with respect to the solution and the 60 nm and 90 nm thick films, with the shift being about $\Delta\lambda \sim 10$ nm. The intensity of this blue shifted band is suppressed rapidly with increasing of the film thickness above 60 nm (Fig. 3, Table 1). This observation suggests that at early stages of the film formation part of the dye molecules are bound in a way typical for H-type aggregation, where the aromatic moieties of the dyes stack on top of each other. In the range of wavelength between $\lambda = 400 - 500$ nm one can distinguish three absorption maxima, with their positions slightly depending on the film thickness, while being not considerably different from the corresponding data in solutions (Table 1). At film thicknesses above ~ 600 nm the substrate was found to have no observable effect on the optical absorption. The difference compared to the reference (solution) spectra is the red shifted absorption around λ = 530 nm in all films grown on FTO glass. Such red shift is typical for dye aggregation and - in particular - J-aggregation, where the dipoles of the dye molecules point to the same direction and the consecutive parallel aromatic moieties are shifted with respect to each other.  It should be noted that this red shifted band is not present in samples fabricated on titanium oxide, suggesting that the dye molecules are not aggregated (compared to glass substrates) forming nearly a monomolecular layer on $TiO_2$ surface. On FTO glass the dye molecules are first chemisorbed on the surface and then - on top of each other: meaning that most of the dye molecules are aggregated between themselves, except the interface layer facing FTO. In $TiO_2$ film the dye forms chemical bond via its carboxylic groups to titanium atoms [15] of the $TiO_2$ lattice. The binding nature of dyes with carboxylic functionality and the $TiO_2$ is complicated and arguable [18, 19]. The detailed analysis of the topic is definitely out of the scope of the present paper.  The red shift of the red-most fluorescein absorption component from λ ≈ 485 nm on FTO to λ ≈ 500 nm on $TiO_2$ grown films can be considered as a fingerprint of covalent binding of the dye in $TiO_2$ based samples.  Such films should probably also contain internal binding sites, which can take a certain amount of the dye from the vapor phase with substantially large portion of the molecules bound into monolayer compared to the FTO based films. The aggregated dyes on $TiO_2$, if present in substantial amounts, should be more loosely bound than on the FTO surface as no substantial spectral shifts are evident.

The best-fit approximation of the absorption peaks correlate with the data for fluorescein in EtOH in concentrations from 0.01 mM to 0.20 mM in aqueous solution. Absorption maxima around λ = 434 nm are related to neutral species of fluorescein. Anion species are responsible for the absorption at λ = 453 nm and λ = 472 nm. The peaks around λ = 490 nm presumably originate from dianion species. As the molar absorption coefficients of the dianion, anion and neutral species of fluorescein are 76900 $M^{-1}cm^{-1}$, 29000 $M^{-1}cm^{-1}$ and 11000 $M^{-1}cm^{-1}$, respectively [20], it is obvious that the molar ratio of the dianions in the studied films is low compared to other species. Corresponding molar ratio of different fluorescein species resembles fluorescein in aqueous solution in pH range 3 - 5 [20]. Peak wavelengths and relative intensities remain qualitatively similar for 40 nm -100 nm thick fluorescein films on glass and on FTO covered glass (not shown).

**Table 1.** Best-fit Gaussian fittings for the absorption spectra in the range from λ = 400 nm to λ = 600 nm of fluorescein dye film of three different thicknesses on FTO-coated glass, 60 nm thick film on $TiO_2$ and in two EtOH solutions of different concentrations. P is the peak position [nm], A is the peak amplitude in arbitrary units and FHMW is given in [nm].

**Figure 3.** Experimental absorption spectra (symbols) in the wavelength range from λ = 400 nm to λ = 600 nm for the dye film of various thicknesses on FTO glass and 60 nm thick on ALD-$TiO_2$ substrate. Absorption of fluorescein in 0.01 mM and 0.20 mM aqueous solution is shown in dark gray dashed and gray dotted lines, respectively. Continuous lines are the Gaussian fittings for the experimental

data given by the symbols of the corresponding color (Table 1).

**3.2 Emission Spectra**

Examples of the emission spectra excited at λ = 458 nm of six fluorescein films of different thickness on FTO covered glass are presented in Fig. 4. As a reference, the emission spectra of 0.20 mM fluorescein in EtOH solution and of dry fluorescein powder are included. For the 0.20 mM EtOH solution the maximum is observed around λ = 517 nm. The peak position is in accordance with the reported emission of fluorescein [20-22]. For the films the principal emission components are located at λ = (520+/-5) nm and λ = (555+/-5) nm. Gaussian fittings resolve the third minor component around λ = (590+/-12) nm. For films thicker than 60 nm an exceptional maxima in the emission spectra is observed at λ ≈ 560 nm. This fairly narrow peak triples its intensity with the increase of the film thickness from 90 nm to 150 nm, and the maxima half width reduced from δλ = 5 nm to δλ = 3 nm as film thickness approaches 150 nm. This peak is not resolved for the 600 nm thick films. With the increase of the film thickness the λ = 520/555 nm doublet clearly looses its intensity due to the aggregation quenching and self absorption. The emission spectra of the studied fluorescein films (excluding the narrow peaks) are qualitatively similar to the data of fluorescein in solution in the anionic form. An emission from dianionic species can be considered to be rather unlikely [20].

**Figure 4.** Experimental emission spectra (symbols) of fluorescein films of different thicknesses on FTO-covered glass: 30 nm (□), 60 nm (◊), 90 nm (▼), 120 nm (▲),

150 nm (○) and 600 nm (□). As a reference the emission spectra at λ = 458 nm excitation of 0.20 mM fluorescein in EtOH ( * ) and dry powder (X) are shown in light gray and gray lines, respectively.

Smooth and highly transparent glass substrates compared to FTO-covered glass provide better sensitivity for the measurement of fluorescein emission starting from the monolayer film thickness. Main emission peaks are located at λ = (520+/-5) nm and λ = (550+/-5) nm (Fig. 5). Thinnest films showed a fine structure absent in thicker samples (Fig. 5, inset). Fitting the fine structure requires involvement of the two narrow δλ =3 nm to δλ =4 nm peaks centered at λ = 540 nm and λ = 560 nm presumably originating from vibrational modes. An interesting feature of the thinnest 1 nm thick film is the absence of the λ = 550 nm band. Appearance of the λ = 550 nm signal at thicknesses starting from ~ 2.5 nm is presumably related to the onset of aggregation of the dye molecules during the island formation stage. Individual islands are disconnected until the film thickness reaches a critical value ~ 10 nm (Fig. 2). Below the 20 nm the emission intensity increases and the spectrum broadens with the increase of the film thickness. Starting from 20 nm thickness up to the thickest studied samples (600 nm) the emission intensity was found to vary very little (Fig. 5).

**Figure 5.** Experimental emission spectra from 1 nm to 600 nm thick fluorescein films on glass substrate measured at λ = 458 nm excitation (symbols). Gaussian fittings are shown with continuous lines. Inset shows the fine structure of the emission spectra for the 1 nm and 2.5 nm thick films.

Fluorescence spectra of the films thicker than 5 nm can be fitted with two major components λ = 520 nm and λ = 550 nm and an additional component at λ = 590 nm, which takes into account the widening of the emission in the red end of the spectrum. Only minor differences were observed between the glass and FTO-covered glass samples. Comparison of the emission spectra of films and fluorescein in EtOH solution suggests anionic-type emission in the former ones [20]. There is a similarity in the emission spectra of the thin fluorescein layers on glass and the thicker layers on FTO-covered glass. For instance, the narrow λ = 560 nm emission peak in 1 nm to 20 nm thick films on glass, and 60 nm to 150 nm samples on FTO covered glass presumably originates from the highly organized fluorescein on the surface (Figs. 3 and 4). In FTO -covered glass samples narrowing of the peak is clearly visible as the film thickness increases that may be related to the reorganization of fluorescein molecules into a highly ordered arrangement [23]. On FTO-covered glass with the RMS surface roughness ~ 20 nm formation of a continuous dye layer requires more than 20 fluorescein monolayers to be grown, whereas on glass substrate a continuous layer is formed already at ~ 10 nm thicknesses (Fig. 2). In case of ALD-$TiO_2$ no fluorescence was observed at excitation λ = 458 nm. The effect could be assigned to the electron injection from the dye molecules into the conduction band of the semiconductor that efficiently quenches emission.

## 4. Conclusions

Vacuum deposition of dye molecules was found to enable highly controllable fabrication of hybrid nanostructures compatible with modern nanolithographic processing. Contrary to conventional thin film preparation from a solution, typically resulting in monolayer thick samples, the thermal evaporation enables fabrication of much thicker films in a controlled manner. Thermal evaporation in vacuum of the studied dye – fluorescein - can be performed efficiently at relatively low temperatures 150 °C - 160 °C (melting point of the dye is 310 °C). Non-oxidizing conditions and low temperature preserve fluorescein from degradation. The deposition conditions can be accurately controlled and monitored so that single monolayer accuracy can be achieved with rather simple arrangement. In practice, when substrates with relatively rough surfaces are used, there is an optimal thickness of dye providing coverage without extensive suppression of the optical emission. Relatively small changes in the deposition parameters may result in pronounced variations of the optical properties of the thin film.

It was found that fluorescein absorption and emission spectra change with the film thickness in a non-trivial manner presumably being related to the thickness dependence of the film morphology. On rough surfaces there is an evidence of ordering of the film structure at thicknesses above several tens of monolayers manifesting itself as narrowing of the $\lambda$ = 560 nm emission peak. On smooth surfaces the corresponding features can be resolved in samples just ten monolayers thick. With the increase of the film thickness the emission is quenched due to the increased molecule aggregation. Quenching of fluorescence of fluorescein on $TiO_2$ substrates is observed in films with thickness ranging from 30 nm to 600 nm. For all substrates the absorption converges to the same spectrum as the film thickness increases over few hundred nm. The optimal thickness for observation of the strong narrow-band emission depends on both the surface roughness and the dye thickness.

The developed method of vacuum-compatible deposition is expected to be applicable to a wide range of organic materials. Preliminary experiments on electron transport properties of vacuum-deposited organic dyes appeared to be promising. Additional studies are necessary to make definite conclusions about applicability of the developed fabrication technique for wider range of optoelectronic applications. If confirmed by further studies, simultaneous evaporation of various dyes and/or transition metal complexes may result in formation of the thin film structures with selective optical properties.

**List of figures and table captions**

**Figure 1.** The evaporator setup. The sample and the crystal oscillator for thickness measurement were assembled symmetrically at the same distance from the crucible.

**Figure 2.** Typical cross sections of the two fluorescein film edges of different thickness on glass substrate measured by atomic force microscope. Insets show the corresponding quasi-3D plots.

**Figure 3.** Experimental absorption spectra (symbols) in the wavelength range from $\lambda$ = 400 nm to $\lambda$ = 600 nm for the dye film of various thicknesses on FTO glass and 60 nm thick on ALD-TiO$_2$ substrate. Absorption of fluorescein in 0.01 mM and 0.20 mM aqueous solution is shown in dark gray dashed and gray dotted lines, respectively. Continuous lines are the Gaussian fittings for the experimental data given by the symbols of the corresponding color (Table 1).

**Figure 4.** Experimental emission spectra (symbols) of fluorescein films of different thicknesses on FTO-covered glass: 30 nm (□), 60 nm (◊), 90 nm (▼), 120 nm (▲), 150 nm (○) and 600 nm (□). As a reference the emission spectra at $\lambda$ = 458 nm excitation of 0.20 mM fluorescein in EtOH ( * )  and dry powder (X) are shown in light gray and gray lines, respectively.

**Figure 5.** Experimental emission spectra from 1 nm to 600 nm thick fluorescein films on glass substrate measured at $\lambda$ = 458 nm excitation (symbols). Gaussian

fittings are shown with continuous lines. Inset shows the fine structure of the emission spectra for the 1 nm and 2.5 nm thick films.

**Table 1.** Best-fit Gaussian fittings for the absorption spectra in the range from λ = 400 nm to λ = 600 nm of fluorescein dye film of three different thicknesses on FTO-coated glass, 60 nm thick film on TiO$_2$ and in two EtOH solutions of different concentrations. P is the peak position [nm], A is the peak amplitude in arbitary units and FHMW is given in [nm].

**Table 1.**

| Sample | P/ A/FHMW | P/ A/FHMW | P/ A/FHMW | P/ A/FHMW |
|---|---|---|---|---|
| 30 nm FTO | 421/ 0.008/ 6 | 462/ 0.049/ 15 | 531/0.046/ 22 | |
| 60 nm FTO | 428/ 0.036/ 13 | 454/ 0.070/ 14 | 484/ 0.062/ 16 | 544/ 0.065/ 24 |
| 90 nm FTO | 428/ 0.027/ 10 | 454/ 0.104/ 19 | 488/ 0.071/ 14 | 545/ 0.070/ 21 |
| 60 nm TiO$_2$ | 427/ 0.002/ 6 | 468/ 0.080/ 31 | 500/ 0.030/ 10 | |
| 0.10$^{-3}$ M | 431/ 0.010/ 16 | 453/ 0.010/ 9 | 483/ 0.025/ 19 | 506/ 0.010/ 9 |
| 0.20$^{-3}$ M | 435/ 0.038/ 17 | 458/ 0.047/ 10 | 486/ 0.063/ 13 | 512/ 0.007/ 10 |

**Figures**

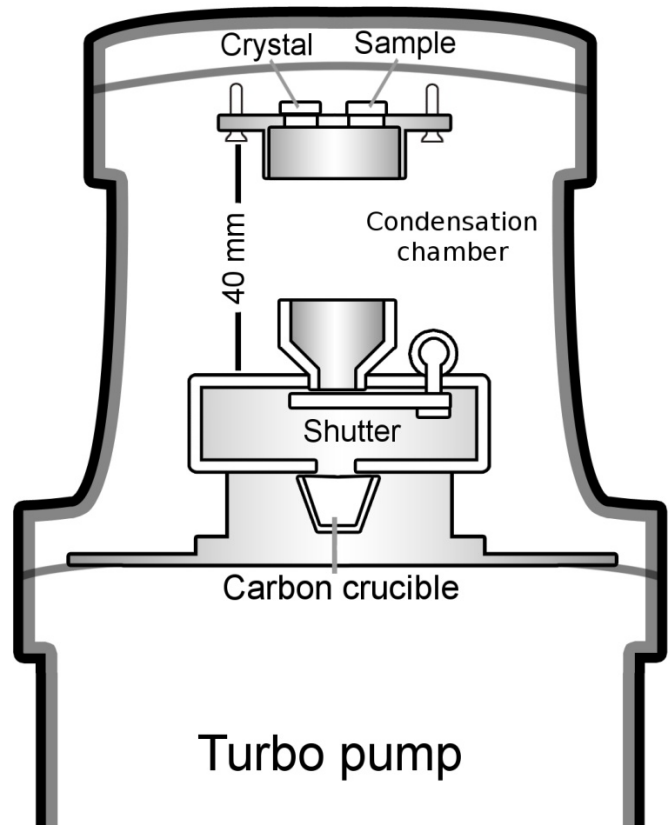

**Figure 1.**

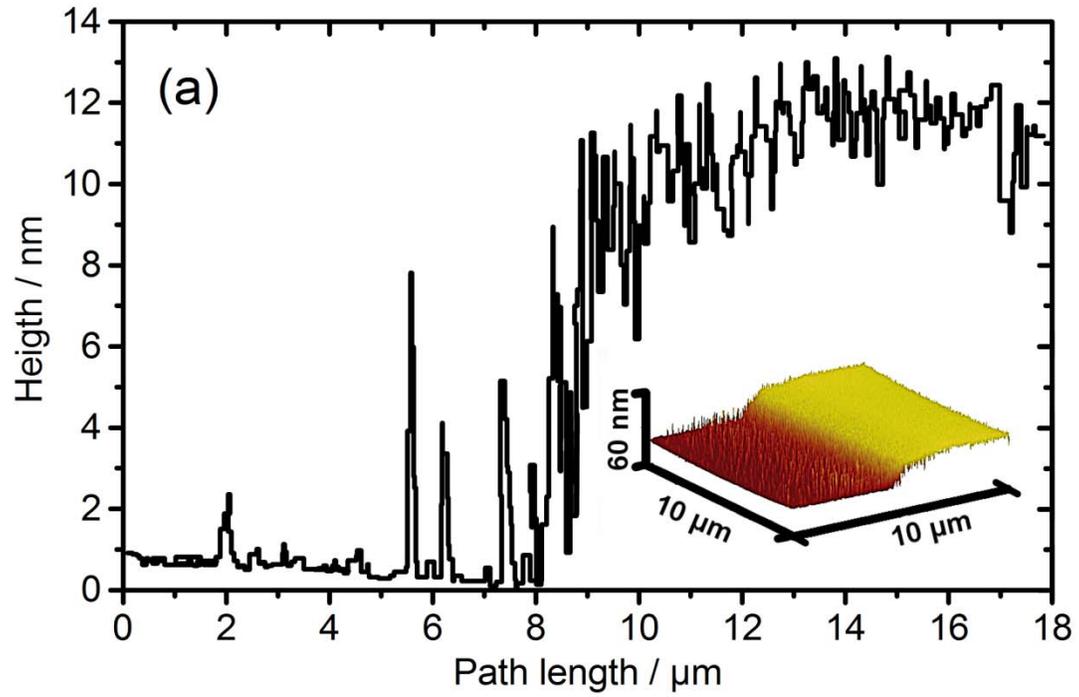

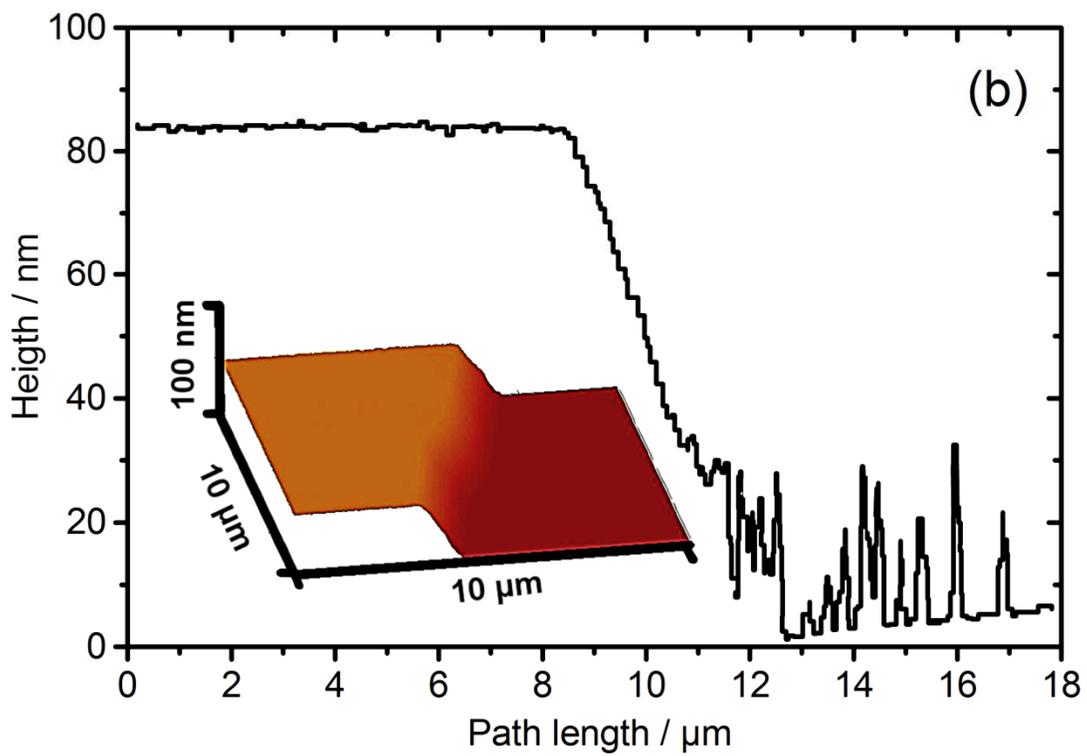

**Figure 2.**

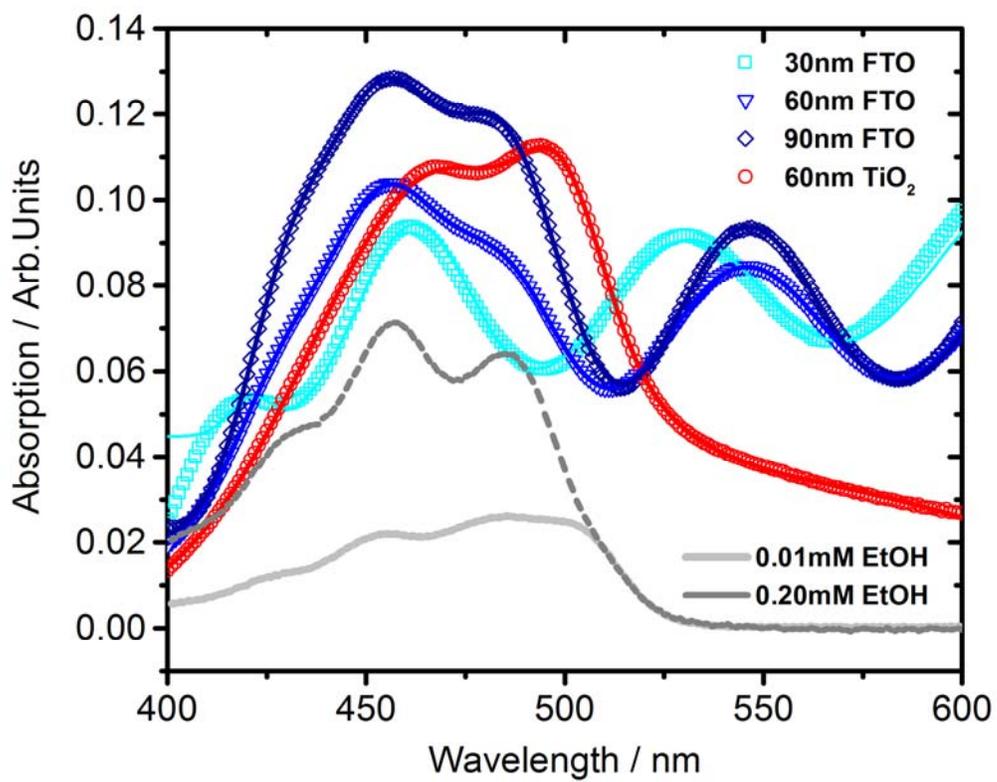

**Figure 3.**

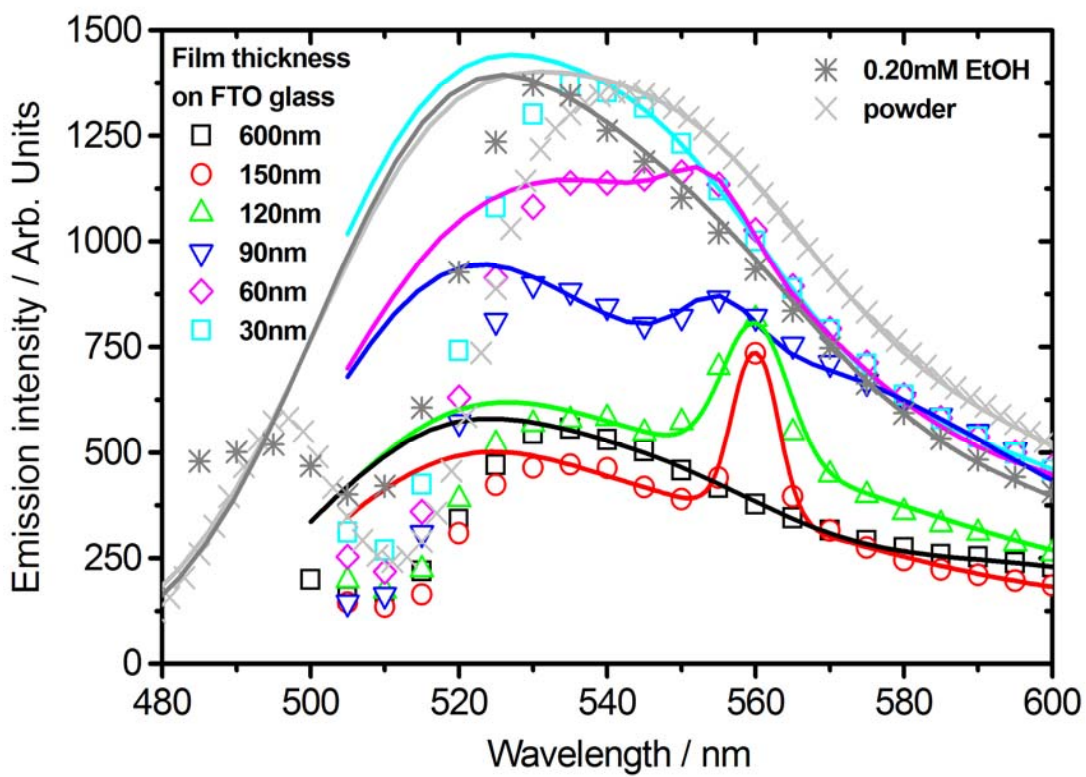

**Figure 4.**

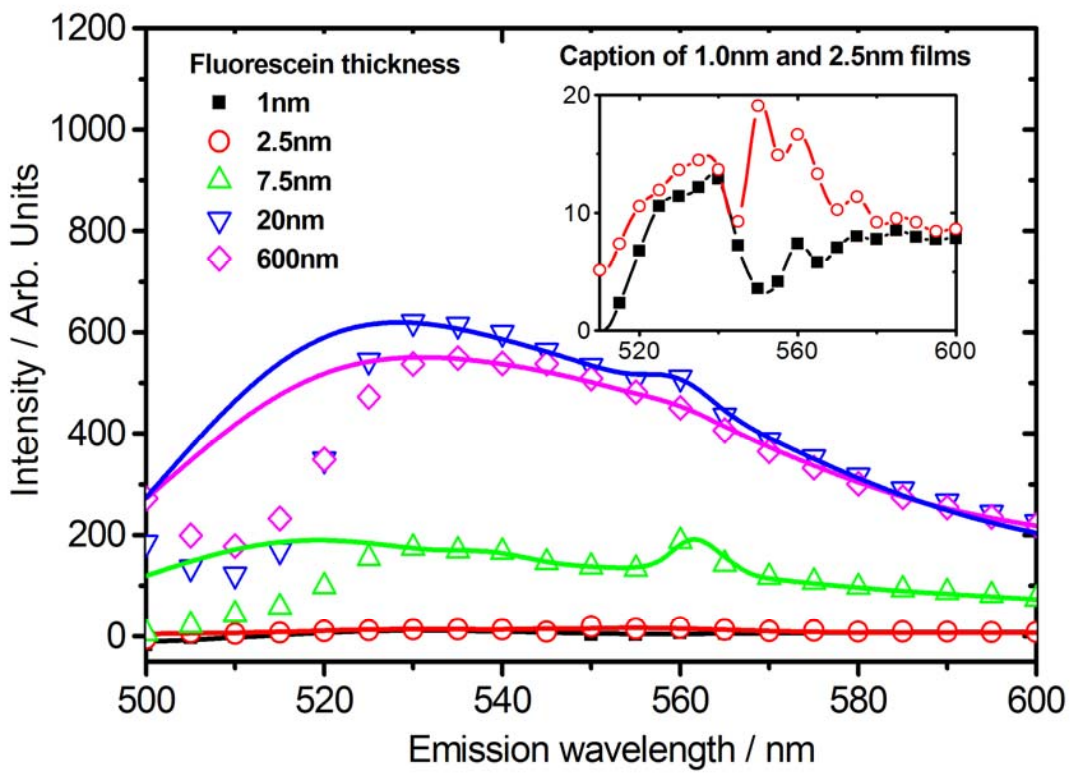

**Figure 5.**